\documentclass[epj]{webofc}
\usepackage[varg]{txfonts}   
\usepackage{graphicx,epsfig,color}
\wocname{EPJ Web of Conferences}
\woctitle{CONF12}
\newcommand{\bc}{\begin{center}}
\newcommand{\ec}{\end{center}}
\newcommand{\be}{\begin{equation}}
\newcommand{\ee}{\end{equation}}
\newcommand{\bea}{\begin{eqnarray}}
\newcommand{\eea}{\end{eqnarray}}
\newcommand{\ba}{\begin{array}}
\newcommand{\ea}{\end{array}}
\newcommand{\lb}{\label}
\newcommand{\rf}{\ref}
\newcommand{\bfg}{\begin{figure}[htbp]}
\newcommand{\efg}{\end{figure}}
\newcommand{\pr}{Phys. Rev. }
\newcommand{\prd}{Phys. Rev. D }

\newcommand{\npb}{Nucl. Phys. B }
\newcommand{\prl}{Phys. Rev. Lett. }
\newcommand{\prp}{Phys. Rep. }
\newcommand{\ap}{Ann. Phys. (N.Y.) }

\newcommand{\plb}{Phys. Lett. B }

\newcommand{\nc}{Nuovo Cimento }

\newcommand{\epj}{Eur. Phys. J. }

\newcommand{\cpc}{Chin. Phys. C }

\begin{document}
\selectlanguage{english}
\title{Introduction to chiral symmetry in QCD}
%
%

\author{H. Sazdjian\inst{1}\fnsep\thanks{\email{sazdjian@ipno.in2p3.fr}} 
}

\institute{Institut de Physique Nucl\'eaire, CNRS-IN2P3,
Universit\'e Paris-Sud, Universit\'e Paris-Saclay,\\
91406 Orsay Cedex, France
}

\abstract{%
The main aspects of chiral symmetry in QCD are presented.
The necessity of its spontaneous breakdown is explained. 
Some low-energy theorems are reviewed. 
The role of chiral effective Lagrangians in the formulation and
realization of chiral perturbation theory is emphasized. 
The consequences of the presence of anomalies are sketched.
}
\maketitle
\section{Introduction} \lb{s1}
Symmetries play an important role in quantum field theory. 
(For general surveys, one may consult, e.g., Refs.
\cite{Coleman:1985,Itzykson:1980rh,Kaku:1993ym,Peskin:1982mu,
Gasser:1982ap,Manohar:1996cq,Scherer:2002tk,Gasser:2003cg,
Ecker:2006xd,Bernard:2006gx,Kubis:2007iy,Brambilla:2014jmp}.)
They introduce limitations in the choice of possible interactions
for a given physical problem or phenomenon and often they
completely fix the structure of the Lagrangian of the theory.
One distinguishes two types of symmetry, local ones, where the
parameters of the transformations are spacetime dependent, and
global ones, where the latter are spacetime independent.
Local symmetries lead in general to the introduction of gauge
theories, while global symmetries classify particles according to
quantum numbers or predict the existence of massless particles.
\par
QCD, the theory of strong interaction, is a gauge theory with the
local symmetry group $SU(N_c)$, acting in the internal space of 
color degrees of freedom. In the real world, $N_c=3$, but leaving
$N_c$ as a free parameter allows one to study the properties of
the theory in more generality. The fundamental fields are the 
quarks (matter fields) and the gluons (gauge fields).
\par
Quark fields, with a common mass parameter $m$,  belong to the 
defining fundamental representation of the color group, which is 
$N_c$-dimensional, antiquark fields to the complex conjugate 
representation of the latter ($N_c$-dimensional), while the gluon 
fields, which are massless, belong to the adjoint representation 
($(N_c^2-1)$-dimensional).
\par     

\section{Flavor symmetry} \lb{s2}

QCD is a confining theory, in the sense that quarks and gluons
are not observed in nature as individual particles. It is admitted 
that only gauge invariant objects should be observable. Hadrons 
are gauge invariant (color singlet) bound states of quarks and gluons
and are observed in nature as free particles.
\par 
From the spectroscopy of hadrons one deduces that there are several 
types of quark, currently six, with different masses, having the
same properties with respect to the gluon fields. One distinguishes 
them with a new quantum number, called flavor. One thus has the 
quarks $u,\ d,\ s$,\ $c,\ b,\ t$.
\par     
In the present study, we are mainly interested in the flavor sector 
of quarks. To simplify notation, we omit, in general, the color indices 
from the quark and gluon fields. The gluons, being flavor singlets, do 
not have flavor indices. In gauge invariant quantities, an implicit 
summation on color indices is understood. Like the number of quark
colors, $N_c$, the number of quark flavors, designated by $N_f$, will 
be considered, for the moment, as a free parameter. 
\par
Let us consider the idealized situation where all $N_f$ quarks have 
the same mass $m$. We assign the quark and antiquark fields to the 
$N_f$-dimensional representation and to its complex conjugate one, 
respectively, with respect to the special unitary group $SU(N_f)$ 
acting in the internal space of flavors. The quark part of the 
Lagrangian density becomes
\be \lb{e2}
\mathcal{L}_q=\overline\psi_a\ \Big(i\gamma^{\mu}
(\partial_{\mu}+igA_{\mu})-m\Big)\ \psi^a.
\ee
(Summation of repeated indices is henceforth assumed: $a$ runs from
1 to $N_f$.)
It is invariant under the continuous global transformations of
the group, with spacetime independent parameters. Designating
the latter by $\alpha^A$ ($A=1,\ldots,N_f^2-1$), one has in 
infinitesimal form
\be \lb{e3}
\delta\psi^a=-i\delta\alpha^A(T^A)_{\ b}^a\psi^b,\ \ \ \ \ 
\delta\overline\psi_a=i\delta\alpha^A\overline\psi_b(T^A)_{\ a}^{b},
\ \ \ \ \delta A_{\mu}=0,
\ee
where $T^A$ are $N_f\times N_f$ hermitian traceless matrices;
they are the representatives of the generators
of the group and hence satisfy the $SU(N_f)$ algebra
\be \lb{e4}
[T^A,T^B]=if_{ABC}T^C,\ \ \ \ \ A,B,C=1,\ldots,N_f^2-1.
\ee
The $f$s are called the {structure constants} of the algebra; they 
are real and completely antisymmetric in their indices.
\par
Using Noether's theorem, one finds $(N_f^2-1)$ {conserved currents}
\be \lb{e5}
j_{\mu}^A(x)=-i\frac{\partial \mathcal{L}_q}
{\partial(\partial^{\mu}\psi^a)}(T^A)_{\ b}^a\psi^b
=\overline\psi_a\gamma_{\mu}(T^A)_{\ b}^a\psi^b,\ \ \ \ \ \ \
\partial^{\mu}j_{\mu}^A=0.
\ee
\par
The generators of the group transformations (also called charges) are
obtained from $j_0^A$ by space integration:
\be \lb{e7}
Q^A=\int d^3x j_0^A(x).
\ee
\par
Because of current conservation, the generators are independent of
time (the fields are assumed to vanish at infinity) and therefore
they commute with the Hamiltonian of the system:
\be \lb{e8}
[H,Q^A]=0.
\ee
They satisfy, as operators, the $SU(N_f)$ algebra
\be \lb{e9}
[Q^A,Q^B]=if_{ABC}Q^C,\ \ \ \ \ A,B,C=1,\ldots,N_f^2-1.
\ee
\par
The system under consideration is thus characterized by the existence
of $N_f$ quark fields with {equal} free masses $m$. Since the 
interaction term itself is invariant under the group transformations, 
the latter property (equality of masses) is preserved after 
renormalization.
\par   
One might also transcribe the transformation properties of the 
fields into similar properties of states.
Ignoring for the moment the confinement problem and introducing 
$N_f$ one-particle quark states $|\mathbf{p}, a>$ (spin labels 
omitted), created from the vacuum state by the fields $\psi^+_a$, 
and assuming the vacuum state is invariant under the transformations 
of the group, one obtains
\be \lb{e10}
Q^A|\mathbf{p}, a>=(T^A)_{\ a}^b|\mathbf{p}, b>.
\ee
Equation (\rf{e8}) implies that the various one-particle states of 
the fundamental representation multiplet have equal masses $m$.
\par
This mode of realization of the symmetry is called the 
\textit{Wigner-Weyl mode}.
\par
Other types of relationship can be found between observables 
involving the multiplet particles, such as form factors or 
scattering amplitudes.
\par

\section{Approximate flavor symmetry with hadrons} \lb{s3}

Consider now the situation where the quarks have different masses,
which ultimately corresponds to the real world. 
The Lagrangian density (\rf{e2}) is replaced by 
\be \lb{e12}
\mathcal{L}_q=\sum_{a=1}^{N_f}\overline\psi_a\ \Big(i\gamma^{\mu}
(\partial_{\mu}+igA_{\mu})-m_a^{}\Big)\ \psi^a.
\ee
The mass term of the Lagrangian density is no longer invariant
under the transformations of the group $SU(N_f)$ and 
therefore the currents obtained previously from Noether's theorem
are not conserved. One finds
\be \lb{e13}
\partial^{\mu}j_{\mu}^A=-i\sum_{a,b=1}^{N_f^{}}(m_a^{}-m_b^{})
\overline\psi_a(T^A)_{\ b}^a\psi^b\neq 0.
\ee
The nonconservation of the currents is thus due to the mass differences
within the representation multiplet.
\par
If, however, the mass differences are much smaller than the 
interaction mass scale, which for QCD is of the order of 
$\Lambda_{QCD}\sim 300$ MeV, or of a multiple of it, 
one is entitled to treat the effects of the mass
differences as perturbations with respect to the symmetric limit
where all the masses are equal. One might write the Lagrangian density 
in the form
\be \lb{e14}
\mathcal{L}=\mathcal{L}_0^{}+\Delta\mathcal{L},
\ee
where $\mathcal{L}_0^{}$ is the Lagrangian density with equal masses 
$m$ and $\Delta\mathcal{L}$ corresponds to mass difference terms.
\par
The above procedure does not destabilize the results of the symmetric 
theory after renormalization. The reason is that mass operators are 
{soft} operators, because their quantized dimension is two for
scalar fields and three for fermion fields, smaller than the dimension
four of the Lagrangian density and in particular of its interaction part.
This has the consequence that mass terms introduce mild effects through
renormalization and at the end the latter remain perturbative.
In particular, they do not affect the currents with anomalous dimensions
\cite{Preparata:1969hg}.
\par
The situation would be different had we introduced the symmetry
breaking through the coupling constants of the interaction terms,
by assigning a different coupling constant to each flavor type quark.
Those have dimension four and their effect on renormalization is 
{hard}. At the end, one generally does not find any trace of 
approximate symmetry, even if at the beginning the coupling constants 
had been only slightly modified.
\par
Since hadrons are bound states of quarks and gluons, 
$SU(N_f)$ flavor symmetry, and more generally its
approximate realization, should be reflected in their properties.
In the exact symmetry case, hadrons should be classified in
$SU(N_f)$ multiplets with degenerate masses.
\par
Considering the real world with six flavor quarks, we observe
that they are divided in two groups: the \textit{light quarks}
$u,\ d,\ s$ and the \textit{heavy quarks} $c,\ b,\ t$. The mass 
differences between the two categories being large ($>\ 1$ GeV), 
an approximate symmetry can be expected only within the space of 
the three light quarks. Therefore, flavor symmetry would be concerned 
either with  $SU(2)$ (isospin symmetry), involving the 
quarks $u$ and $d$, or with $SU(3)$, involving the quarks 
$u,\ d,\ s$.
\par 
Concerning isospin symmetry, one notes the approximate equalities of 
the masses of the proton and the neutron, of the charged and neutral 
pions, of the kaons and of many other groups of particles.
The nucleons and the kaons could be placed in doublet (fundamental)
representations of $SU(2)$ (isospin 1/2), the pions in 
the triplet (adjoint) representation (isospin 1), the 
$\Delta$s in the quadruplet representation (isospin 3/2), etc.
\par
Since the mass differences within each multiplet are very small,
of the order of a few MeV, one deduces that the difference
between the masses of the quarks $u$ and $d$ is also of the same 
order:
\be \lb{e15}
|m_d-m_u|\sim \mathrm{a\ few\ MeV}.
\ee
(Precise calculations should also include the contributions of the
electromagnetic interaction, which also are of the same order.)
\par
For the realization of $SU(3)$ symmetry, the quarks $u,\ d,\ s$ are 
assigned to the defining representation $3$, while their 
antiparticles to the representation $\overline 3$. 
\par
Since mesons are made from one quark and one antiquark fields, they
would be classified through the product representation  
\be \lb{e16}
3\otimes\overline{3}\ =\ 8\ \oplus\ 1,
\ee
where $8$ is the octet representation and $1$ the singlet one. On 
phenomenological grounds, one notices that for example the vector 
mesons $\rho$, $K^*$, $\omega$, $\phi$ can be grouped in an octet
plus a singlet and similarly for the pseudoscalar mesons $\pi$, 
$K$, $\eta$, $\eta'$, although for them the mass differences are 
much larger than for the vector mesons. (The understanding of 
the latter phenomenon comes with chiral symmetry).
\par
Baryons, being made from three quark fields, would be classified 
through the product representation
\be \lb{e17}
3\otimes 3\otimes 3\ =\ 10\ \oplus\ 8\ \oplus 8\ \oplus 1.
\ee
This is the case for example of the baryons $N$, $\Sigma$, $\Xi$ and 
$\Lambda$, which may be classified in an octet, of the baryons 
$\Delta$, $\Sigma'$, $\Xi'$ and $\Omega$, which may be classified in 
a decuplet, etc.
\par
The approximate $SU(3)$ symmetry leads to many relations for the mass 
differences of a given multiplet ({Gell-Mann--Okubo formulas}), for the 
coupling constants of particles belonging to multiplets, for form 
factors, etc.
\par
The mass differences within an $SU(3)$ multiplet being of the order 
of 100 MeV, one deduces that the mass difference between the quark 
$s$ and the quarks $u$ and $d$ is of the same order of magnitude and 
much greater than the mass difference between $u$ and $d$:  
\be \lb{e18}
\big(m_s-(m_u+m_d)/2\big)\sim 100\ \mathrm{MeV}\ \gg |(m_d-m_u)|
\sim \mathrm{a\ few\ MeV}.
\ee
\par
The approximate $SU(3)$ symmetry can also be used for
hadrons containing heavy quarks; these, however, should stand as
backgrounds for the group analysis.
For example one could apply the analysis to mesons made of
$\overline q_hq_{\ell}$, or to baryons made of $q_hq_{\ell}q_{\ell}$, 
where $q_h=c,\ b,\ t$ and $q_{\ell}=u,\ d,\ s$.
\par

\section{Chiral symmetry} \lb{s4}

Fermion fields may undergo other types of unitary transformation than 
those met with the flavor symmetry. They are generated with the inclusion 
of the matrix $\gamma_5^{}$ in the former transformations and are called 
\textit{axial flavor} transformations, since they change the parity 
properties of fields. In infinitesimal form they are of the type
\be \lb{e19}
\delta\psi^a=-i\delta\alpha^A(T^A)_{\ b}^a\gamma_5^{}\psi^b,
\ \ \ \ \ 
\delta\overline\psi_a=-i\delta\alpha^A\overline\psi_b\gamma_5^{}
(T^A)_{\ a}^b,\ \ \ \ \delta A_{\mu}=0,
\ee
where the indices $a,\ b,\ A$ refer to the flavor group $SU(N_f)$ 
representations met before and the $T$s are hermitian traceless 
matrices representing the corresponding generators in the fundamental 
representation. 
\par    
We consider the quark part of the QCD Lagrangian density with 
equal mass quarks [Eq. (\rf{e2})].
Under the above transformations, this Lagrangian density is not
invariant, because of the presence of the mass terms:
\be \lb{e20}
\delta\mathcal{L}_q=2im\delta\alpha^A\overline\psi_a
(T^A)_{\ b}^a\gamma_5^{}\psi^b.
\ee
{Thus, invariance under axial flavor transformations requires 
vanishing of the quark mass terms.} Contrary to the ordinary flavor 
symmetry transformations, equality of masses is no longer sufficient 
for ensuring invariance.
\par 
In the more general case of unequal masses, the mass terms can be
represented in the form of a {diagonal matrix} $\mathcal{M}$, such 
that $\mathcal{M}=\mathrm{diag}(m_1^{},m_2^{},\ldots,m_{N_f}^{})$. 
In that case the variation of the Lagrangian density is
\be \lb{e21}
\delta\mathcal{L}_q=i\delta\alpha^A\overline\psi_a
\{\mathcal{M},T^A\}_{\ b}^a\gamma_5\psi^b,
\ee
where $\{,\}$ is the anticommutator.
\par
We now consider the case of massless quarks. The Lagrangian density 
is invariant under both the flavor and axial flavor transformations. 
The conserved currents are
\bea
\lb{e22}
& &j_{\mu}^A(x)=\overline\psi_a\gamma_{\mu}(T^A)_{\ b}^a\psi^b,
\ \ \ \ \ \ \ \ \partial^{\mu}j_{\mu}^A=0, \\
\lb{e23}
& &j_{5\mu}^A(x)=\overline\psi_a\gamma_{\mu}\gamma_5^{}(T^A)_{\ b}^a
\psi^b,\ \ \ \ \ \ \ \ \partial^{\mu}j_{5\mu}^A=0. 
\eea
\par
The corresponding charges are defined from space integration on the 
current densities:
\be \lb{e24}
Q^A=\int d^3x j_0^A(x),\ \ \ \ \ \ Q_5^A=\int d^3x j_{50}^A(x). 
\ee
\par
The flavor and axial flavor transformations form the set of
\textit{chiral} transformations. The corresponding charges satisfy
the following algebra:
\bea \lb{e25}
& &[Q^A,Q^B]=if_{ABC}Q^C,\ \ \ \ \ [Q^A,Q_5^B]=if_{ABC}Q_5^C,
\nonumber \\
& &[Q_5^A,Q_5^B]=if_{ABC}Q^C,\ \ \ \ \ A,B,C=1,\ldots,N_f^2-1.
\eea
Note that the axial charges do not form alone an algebra. The previous
algebra can, however, be simplified and become more transparent.
Define
\be \lb{e26}
Q_L^A=\frac{1}{2}(Q^A-Q_5^A),\ \ \ \ \ \ \ Q_R^A=\frac{1}{2}(Q^A+Q_5^A).
\ee
($L$ for left-handed and $R$ for right-handed.) One obtains
\be \lb{e27}
[Q_L^A,Q_L^B]=if_{ABC}Q_L^C,\ \ \ \ \ \ \ [Q_R^A,Q_R^B]=if_{ABC}Q_R^C,
\ \ \ \ \ \ \ \ [Q_L^A,Q_R^B]=0.
\ee
Therefore, the left-handed and right-handed charges are decoupled
and operate separately. Each of them generate an $SU(N_f)$
group of transformations. The whole chiral group is then decomposed
into the direct product of two $SU(N_f)$ groups, which will
be labeled with the subscripts $L$ and $R$, respectively: 
\be \lb{e28}
\mathrm{Chiral\ group}\ =\ SU(N_f)_L^{}\ \otimes\ SU(N_f)_R^{}. 
\ee
\par
The ordinary flavor transformations form a subgroup of these,
denoted $SU(N_f)_V^{}$:
\be \lb{e29}
\mathrm{Flavor\ group}\ =\ SU(N_f)_V^{}.
\ee
(The subscript $V$ refers to the vector nature of the corresponding 
currents.)
\par

\section{Explicit chiral symmetry breaking} \lb{s5}

In nature, quarks have masses. Therefore, chiral symmetry cannot
be an exact symmetry of the QCD Lagrangian. The quark mass terms 
introduce an {explicit} chiral symmetry breaking. 
\par
The symmetry breaking could be treated as a perturbation only if
the quark masses are much smaller than the QCD mass scale. This
eliminates the heavy quarks $c,\ b,\ t$ from the domain of 
investigations. We are left with the sector of light quarks 
$u,\ d,\ s$ and the {approximate chiral symmetry} 
$SU(3)_L\otimes SU(3)_R$.
\par  
What would be the signature of this approximate symmetry in 
nature?
The axial charges $Q_5^A$ are pseudoscalar objects. When acting, 
in the chiral symmetry limit (massless quarks) and in the Wigner-Weyl 
mode, on a massive hadronic state belonging to a multiplet of the 
flavor group, they would produce new states with the same mass and spin, 
but with opposite parity:
\be \lb{e30}  
Q_5^A\ |M,s,\mathbf{p},+,a>=(T^{\prime A})_{\ a}^b|M,s,\mathbf{p},-,b>.
\ee
The $+$ and $-$ labels refer to the intrinsic parity of the states; 
for definiteness, we have assumed above that the initial state has 
positive parity. (For chiral singlet representations the matrix 
$T'$ is null.)
\par
Thus, if we adopt the Wigner-Weyl mode of realization of chiral 
symmetry, we should find \textit{parity doublets} for most of 
massive states. 
\par
When the light quarks obtain masses, the degeneracy of masses within 
the parity doublets would be removed, but the masses would still 
remain close to each other.
However, no parity doublets of massive particles with approximately
equal masses are found, in general, in the hadronic world.
This observation forces us to abandon the Wigner-Weyl mode of
realization of chiral symmetry.
\par
The other alternative that remains is the \textit{phenomenon of 
spontaneous symmetry breaking}, also called the 
\textit{Nambu-Goldstone mode} of realization of chiral symmetry
\cite{Goldstone:1961eq,Goldstone:1962es,Nambu:1960xd,Nambu:1961tp}.
\par

\section{Spontaneous chiral symmetry breaking} \lb{s6}

An inherent assumption within the Wigner-Weyl mode is that the
ground state of the theory (the {vacuum state} in QFT) is 
{invariant} under the symmetry group of transformations.
This means that the generators of the transformations
annihilate the vacuum state:
\be \lb{e31}
Q^A\ |0>=0.
\ee
(Here, the specific case of vector flavor charges has been considered.)
\par
Since one-particle states are constructed from the action of the
fields on the vacuum state, the above property guarantees that 
one-particle states do transform as elements of irreducible 
representations of the symmetry group. 
\par
However, the vacuum state (the ground state) may not be invariant
under the symmetry group of transformations, even if the Lagrangian
is. In that case, it is not a symmetric state and the generators of 
the symmetry group do not annihilate it:
\be \lb{e32}
Q_5^A\ |0>\neq 0.
\ee
(Here, the specific case of axial-vector flavor charges has been 
considered.) 
It is said that the symmetry is \textit{spontaneously broken}.
\par
A similar situation is found in the well-known {sigma model} 
\cite{GellMann:1960np} or in the $O(N)$ model of spin-0 fields 
\cite{Manohar:1996cq}. 
Here, one studies the properties of the potential energy at the classical 
level. According to the values of the parameters of theory, the potential 
energy may have a non-symmetric ground state.
\par   
The absence of parity doublets for massive hadronic states forces us
to explore the possibility of spontaneous breaking of chiral symmetry.
The axial charges, when applied on the vacuum state, would now produce
new states:
\be \lb{e33}
Q_5^A\ |0>=|\chi^A, ->,\ \ \ \ \ \ A=1,...,8.
\ee
\par
These states have the same quantum properties as their generating
axial charges, the vacuum state being assumed to have a positive
parity and no quantum numbers. In particular, they are pseudoscalar 
states. Since in the symmetric limit the charges commute with the 
Hamiltonian, the energy of the states $|\chi^A,->$ is null. This is 
possible only if there exist \textit{massless pseudoscalar particles}, 
which might also generate by superposition other many-particle 
zero-energy states. This is the content of the \textit{Goldstone theorem}
\cite{Goldstone:1961eq,Goldstone:1962es}.
We note, however, that the states $|\chi^A,->$ are not one-particle 
states, but rather a superposition of many-massless-particle states;
their norm is infinite. The fact that these states do not form 
new degenerate vacua, but rather represent zero-energy limits of
many-massless-particle states, hinges on the assumption that the
hadronic vacuum state $|0>$ is unique. This is corroborated by the
fact that when the quarks obtain small masses, explicitly breaking 
chiral symmetry, these particles become massive, although with small
masses, while the state $|0>$ evolves towards the true vacuum state
of the theory.   
\par   
Spontaneous chiral symmetry breaking is thus manifested by the
existence of eight pseudoscalar massless particles (mesons), called
\textit{Nambu-Goldstone bosons}. 
\par
Spontaneous breaking concerns, however, only the axial sector of the
charges. The ordinary flavor symmetry is still realized with the 
Wigner-Weyl mode. Therefore the symmetry group
$SU(3)_L^{}\otimes SU(3)_R^{}$ is spontaneously broken 
down to the flavor group $SU(3)_V^{}$:
\be \lb{e34}
SU(3)_L^{}\otimes SU(3)_R^{}\ \longrightarrow\ SU(3)_V^{}.
\ee
\par
In the real world, where quarks have masses, chiral symmetry will
undergo an additional explicit symmetry breaking. Under this effect,
the eight Nambu-Goldstone bosons will acquire small masses, as compared
to the masses of massive hadrons.
\par
Considering the spectroscopy of mesons, one notices the existence
of eight light pseudoscalar mesons, $\pi,\ K,\ \eta$. They can be 
identified, in the chiral limit, with the eight Nambu-Goldstone 
bosons. 
\par
One would expect that the masses-squared of these particles are 
proportional to the masses of the quarks that make their content, 
which would explain in turn the mass differences and hierarchies 
between them. Thus, one should have (cf. Sect. \rf{ss86}) 
\be \lb{e35}
M_P^2=O(\mathcal{M}),
\ee
while for the other hadronic massive states, one should have 
the expansion
\be \lb{e36}
M_h^2\ =\ M_{h0}^2\ +\ O(\mathcal{M}),
\ee
$M_{h0}^2$ being the same for all members of a flavor
multiplet and different from zero in the chiral limit. ($M_{h0}^2$ 
differs from one flavor multiplet to another.)
This explains why the lightest pseudoscalar meson masses are more
sensitive to the quark masses than those of the massive hadrons.
\par

\section{Properties of Goldstone bosons} \lb{s7}

Consider, in the chiral limit, the coupling of the Goldstone
bosons to the axial-vector currents: 
\be \lb{e37}
<0|\ j_{5\mu}^A\ |P^B,p>\ =\ i\delta_{AB}p_{\mu}^{}F_{P0}.
\ee
The coupling $F_{P0}$ is different from zero, since otherwise the axial 
charges, which are constructed from $j_{50}^A$, could not create from 
the vacuum state zero-energy states. 
Using conservation of the axial vector currents and the operator
equation $\partial^{\mu}j_{5\mu}^A=i[P_{\mu},j_{5\mu}^A]$, one
obtains
\be \lb{e38}
<0|\ \partial^{\mu}j_{5\mu}^A\ |P^B,p>\ =\ 0\ =\ 
\delta_{AB}p^2F_{P0}\ =\ \delta_{AB}M_{P0}^2F_{P0},
\ee
which implies that $M_{P0}=0$. This is another verification of the 
Goldstone theorem.
\par
Consider now the coupling of a {massive} pseudoscalar state
(for example a radial excitation of the Goldstone boson)
to the axial-vector current:
\be \lb{e39}
<0|\ j_{5\mu}^A\ |P^{\prime B},p>\ =\ i\delta_{AB}p_{\mu}^{}
F_{P'0}.
\ee
Using conservation of the axial-vector currents one again finds 
\be \lb{e40}
<0|\ \partial^{\mu}j_{5\mu}^A\ |P^{\prime B},p>\ =\ 0\ =\ 
\delta_{AB}p^2F_{P'0}\ =\ \delta_{AB}M_{P'0}^2F_{P'0}.
\ee
However, since $M_{P'0}\neq 0$, one deduces that $F_{P'0}=0$.
This means that the massive pseudoscalar states {decouple} from 
the axial-vector currents.
\par
The above results can be summarized as follows:
\bea 
\lb{e41}
& &\mathrm{Goldstone\ bosons}:\ \ \ M_{P0}^2=0,\ \ \ F_{P0}\neq 0.\\
\lb{e42}
& &\mathrm{Massive\ pseudoscalar\ mesons}:\ \ \ M_{P'0}^2\neq 0,\ \ \  
F_{P'0}=0.
\eea
\par
When quarks obtain masses, the above properties are modified 
by terms proportional to the quark masses.
\bea 
\lb{e43}
& &\mathrm{Goldstone\ bosons}:\ \ \ M_P^2\ =\ O(\mathcal{M}),
\ \ \ \ \ \ F_P^{}\ =\ F_{P0}^{}\ +\ O(\mathcal{M}),\\
\lb{e44}
& &\mathrm{Massive\ pseudoscalar\ mesons}: 
M_{P'}^2\ =\ M_{P'0}^2\ +\ O(\mathcal{M}),\ \ \ \ \ \ 
F_{P'}^{}\ =\ O(\mathcal{M}),\\
\lb{e45}
& &M_P^2\ll M_{P'}^2,\ \ \ \ \ \ \ \ |F_{P'}|\ll |F_{P}|.
\eea
\par   
On experimental grounds, from the leptonic decays of $\pi$ and 
$K$ mesons, one has \cite{Agashe:2014kda}
\be \lb{e46}
F_{\pi}\simeq 92\ \mathrm{MeV},\ \ \ \ \ \ F_K\simeq 110\ \mathrm{MeV}.
\ee
The quantity $(F_K/F_{\pi}-1)\simeq 0.2$ measures the 
order of magnitude of flavor $SU(3)$ breaking.
\par

\section{Low-energy theorems} \lb{s8}

The decoupling of the massive pseudoscalar mesons from the axial-vector 
currents in the chiral limit (massless quarks) allows one to derive 
low-energy theorems concerning processes where enters at least 
one Goldstone boson. Most of these relations are obtained with the
aid of the Ward-Takahashi identities.
\par
Contrary to to the ordinary flavor symmetry, chiral symmetry does not
yield linear relations between matrix elements of multiplets, but
rather leads to relations between processes involving absorption and/or 
emission of Goldstone bosons at low momenta, for example between
the process $\alpha\ \rightarrow\ \beta$ and the process
$\alpha\ +\ n_1P\ \rightarrow\ \beta\ +\ n_2P$, where $n_1$ Goldstone
bosons are absorbed and $n_2$ Goldstone bosons are emitted 
($n_1\ge 0$, $n_2\ge 0$). 
\par
\textit{Remark.} In QCD, the quark masses appear as free parameters.
Therefore, one expects that all hadronic physical quantities -- masses, 
decay couplings, coupling constants, form factors, scattering 
amplitudes -- possess analyticity properties in them, up to the 
existence of cuts or branching points.
These objects appear in general as residues of Green's functions
at physical particle poles. Therefore, they define 
\textit{on-mass shell} quantities. They should not be considered 
as functions of the mass-shell variables $p^2,\ p^{\prime 2}$, 
etc., but only of the quark mass parameters and of the momentum 
transfers or of the Mandelstam variables $s,\ t,\ u$, etc., which,
eventually may take unphysical values by analytic continuation.
Green's functions, on the other hand, may be functions of the 
mass-shell variables $p^2,\ p^{\prime 2}$, etc.
\par

\subsection{Goldberger-Treiman relation} \lb{ss81}

Consider, in the isospin limit, the matrix element of the axial-vector 
current between proton and neutron states \cite{Goldberger:1958vp}:
\be \lb{e47}
<p(p')|\ j_{5\mu}^{1+i2}\ |n(p)>=\overline u_p(p')\ \Big[\ 
\gamma_{\mu}\gamma_5^{}g_A^{}(q^2)+q_{\mu}\gamma_5^{}h_A^{}(q^2)\ 
\Big]\ u_n(p),
\ee
where $q=(p-p')$; $g_A^{}$ and $h_A^{}$ are the axial-vector form 
factors of the nucleons. 
\par
Taking the divergence of the current, one obtains
\be \lb{e48}
<p(p')|\ \partial^{\mu} j_{5\mu}^{1+i2}\ |n(p)>=
-i\Big(2M_N^{}g_A^{}(q^2)+q^2h_A^{}(q^2)\Big)\ 
\overline u_p(p')\gamma_5^{}u_n(p).
\ee
The left-hand side has singularities through the contribution of 
pseudoscalar intermediate states. 
For simplicity and illustrative purposes, let us assume that the
latter can be saturated by a series of narrow-width particles,
composed of the pion ($\pi$) and of its radial excitations
($\pi^n$, $n=1,2,\ldots$). It can be shown 
that multipion states, because of phase space, do not contribute 
to the final result. One obtains 
\be \lb{e49}
<p(p')|\ \partial^{\mu} j_{5\mu}^{1+i2}\ |n(p)>=-2i\Big\{
\frac{M_{\pi}^2F_{\pi}}{M_{\pi}^2-q^2}g_{\pi NN}
+\sum_{n=1}^{\infty}\frac{M_{\pi^n}^2F_{\pi^n}}
{M_{\pi^n}^2-q^2}g_{\pi^n NN}\Big\}\ \overline u_p(p')\gamma_5u_n(p),
\ee
where $g_{\pi NN}^{}$ and $g_{\pi^n NN}^{}$ are the coupling constants 
of the pseudoscalar mesons with the nucleons. Equation (\rf{e48})
becomes
\be \lb{e50}
2\frac{M_{\pi}^2F_{\pi}}{M_{\pi}^2-q^2}g_{\pi NN}
+2\sum_{n=1}^{\infty}\frac{M_{\pi^n}^2F_{\pi^n}}
{M_{\pi^n}^2-q^2}g_{\pi^n NN}\ =\ 2M_N^{}g_A^{}(q^2)+q^2h_A^{}(q^2).
\ee
We take the limit $q^2=0$. In the right-hand side, 
$h_A(q^2)$ does not have a pole at this value (no massless pseudoscalars 
in the real world). This gives 
\be \lb{e51}
F_{\pi}g_{\pi NN}+\sum_{n=1}^{\infty}F_{\pi^n}g_{\pi^n NN}\ 
=\ M_N^{}g_A^{}(0).
\ee
Consider now the $SU(2)_L^{}\otimes SU(2)_R^{}$ chiral limit 
(massless $u$ and $d$ quarks). According to Eq. (\rf{e42}),
all massive pseudoscalar mesons decouple from the axial-vector current. 
Equation (\rf{e51}) reduces to the relation
\be \lb{e52}
g_A^{}(0)\ =\ \frac{F_{\pi}g_{\pi NN}}{M_N^{}}.
\ee
This is an exact result of QCD in the $SU(2)_L^{}\otimes SU(2)_R^{}$
chiral limit (a low-energy theorem). Note that it is independent
of the order of the two limits that were taken. Had we first 
considered the chiral limit, then the left-hand side of Eq. 
(\rf{e48}) would vanish, but at the same time in the right-hand 
side the form factor $h_A^{}(q^2)$ would display a pion pole at 
the position $q^2=0$ and would contribute to the equation. 
\par
The experimental values are: $g_A^{}\simeq 1.27$ \cite{Agashe:2014kda}, 
$g_{\pi NN}^{}\simeq 13.15$ \cite{Baru:2010xn}, $F_{\pi}^{}\simeq 92.2$ 
MeV \cite{Agashe:2014kda}, $M_N^{}=(M_p+M_n)/2=938.92$ MeV. The 
right-hand side is then $\simeq 1.29$, to be compared with the left-hand 
side, $1.27$. The discrepancy is about $2\%$, which is typical of 
the corrections coming from explicit breaking  
of $SU(2)_L^{}\otimes SU(2)_R^{}$ symmetry 
($M_{\pi}^2/M_N^2\simeq 0.02)$.
\par       

\subsection{Ward-Takahashi identities} \lb{ss82}

Ward-Takahashi identities (WTI) \cite{Ward:1950xp,Takahashi:1957xn} are 
obtained by considering correlation functions of axial-vector currents 
with local operators $O(x)$:
\be \lb{e53}
\int dx e^{iq.x} <\beta\ |Tj_{5\mu}^A(x)O(0)|\ \alpha>,
\ee
where $|\ \alpha>$ and $|\ \beta>$ are hadronic states.
\par
Taking the divergence of the axial-vector current gives 
\bea \lb{e54}
& &-iq^{\mu}\int dx e^{iq.x} <\beta\ |Tj_{5\mu}^A(x)O(0)|\ \alpha>\
=\ \int dx e^{iq.x} 
<\beta\ |T\partial^{\mu}j_{5\mu}^A(x)O(0)|\ \alpha>\nonumber \\
& &\ \ \ \ \ \ \ \ \ \ \ +\
\int dx e^{iq.x} \delta(x^0) <\beta\ |[j_{50}^A(x),O(0)]|\ \alpha>.
\eea
One notices the presence of the equal-time commutator, which should
be evaluated in the theory. Because of causality, it should
involve a linear combination of $\delta^3(\mathbf{x})$ and a 
finite number of its derivatives.
\par
One then takes the limit of low or zero values of $q$
and proceeds with similar methods as in the Goldberger-Treiman case.
\par
The method can also be generalized by considering correlation functions
of the axial-vector currents with a multiple number of local operators.
\par

\subsection{Callan-Treiman relation} \lb{ss83}

Choose $|\ \alpha>=|\ K^+>$, $|\ \beta>=|0>$,
$O=j_{\nu}^{4-i5}$ and $j_{5\mu}^A=j_{5\mu}^{3}$ \cite{Callan:1966hu}. 
The equal-time commutator of the WTI yields the axial-vector current 
$j_{5\nu}^{4-i5}$ plus Schwinger terms that do not contribute. 
The matrix element involving the divergence of the 
axial-vector current is again saturated by the pion and its radial
excitations. The corresponding residues are proportional to the
$K_{\ell 3}$ form factors with respect to the pion and
to its radial excitations.
\par
One has the definition 
\be \lb{e55}
<\pi^0(p')\ |\ j_{\nu}^{4-i5}\ |\ K^+(p)>=\frac{1}{\sqrt{2}}
\Big[\ (p+p')_{\nu}^{}f_+(t)+(p-p')_{\nu}^{}f_-(t)\ \Big],
\ \ \ \ \ t=(p-p')^2,
\ee
with similar definitions for the radial excitations of the pion.
Take the limit $q\rightarrow 0$ in the WTI [Eq. (\rf{e54})], in which 
case the left-hand side vanishes (no poles at $q=0$). 
Then the $SU(2)_L\otimes SU(2)_R$ chiral limit is taken 
(massless $u$ and $d$ quarks). The massive pseudoscalar states 
decouple and one ends up with the relation  
\be \lb{e56}
f_+(M_K^2)+f_-(M_K^2)=\frac{F_K^{}}{F_{\pi}^{}}.
\ee
\par
The physical domain of the $K_{\ell 3}$ form factors
(corresponding to the decay $K\rightarrow \pi\ell\nu$)
being limited by the inequalities 
$m_{\ell}^2\leq t\leq (M_K^{}-M_{\pi}^{})^2$, the form
factors appearing in the above relation are evaluated at the 
unphysical point $t=M_K^2$. Extrapolations are used from the physical 
domain to reach that point. The relation is well satisfied on 
experimental grounds with a few percent of discrepancy.
\par  
The Callan-Treiman formula establishes a relation between the form 
factors of the process $K\rightarrow \pi\ell\nu$ and
the decay coupling of the process $K\rightarrow \ell\nu$.
\par

\subsection{Pion scattering lengths (Weinberg)} \lb{ss84}

Choose for $|\ \alpha>$ and $|\ \beta>$ in the WTI [Eq. (\rf{e54})]  
target particles like $N$, $K$ or $\pi$, and for $O$ and $j_{5\mu}^A$ 
axial vector currents with the pion quantum numbers 
\cite{Weinberg:1966kf}.
Then calculate the divergences of the two currents. The WTI involves
now two momenta $q$ and $k$ and two equal time commutators. The 
momenta squared $q^2$ and $k^2$ are taken to zero, but $q$ and $k$ 
are maintained nonzero, of the order of $M_{\pi}$. 
One ends up with formulas for the S-wave scattering lengths of the 
processes $\pi+\alpha\rightarrow \pi+\alpha$, where  
$\alpha$ is the target particle, much heavier than the 
pion:
\be \lb{e57}
a_0^I=-\frac{M_{\pi}}{8\pi F_{\pi}^2}(1+\frac{M_{\pi}}{M_{\alpha}})^{-1}
\ [I(I+1)-I_{\alpha}(I_{\alpha}+1)-2],
\ee
where $I$ is the total isospin of the state $|\ \pi\ \alpha>$ and 
$I_{\alpha}$ the isospin of the target particle.
This formula is applied to the scattering processes
$\pi+N\rightarrow \pi+N$ and $\pi+K\rightarrow \pi+K$
\cite{Weinberg:1966kf,Tomozawa:1966jm}.
\par
When the target particle is the pion itself, the analysis should be 
completed by retaining higher-order terms in the kinematic variables. 
Crossing symmetry is also used. One then obtains the $\pi-\pi$ 
scattering amplitude at low energies:
\be \lb{e58}
\mathcal{M}_{ac,bd}=\frac{1}{F_{\pi}^2}\{\ \delta_{ac}\delta_{bd}
(s-M_{\pi}^2)+\delta_{ab}\delta_{cd}(t-M_{\pi}^2)
+\delta_{ad}\delta_{bc}(u-M_{\pi}^2)\ \},
\ee
where $a,b,c,d$ are the pion isospin indices and 
$s,t,u$ the Mandelstam variables.
\par
The S-wave scattering lengths are
\be \lb{e59}
a_0^0=\frac{7M_{\pi}}{32\pi F_{\pi}^2}\simeq 0.16\ M_{\pi}^{-1}, 
\ \ \ \ \ \ 
a_0^2=-\frac{2M_{\pi}}{32\pi F_{\pi}^2}\simeq -0.046\ M_{\pi}^{-1}. 
\ee
The above predictions are well satisfied experimentally within
$10-25\%$ of discrepancy. Direct measurements of the scattering 
lengths are, however, not possible because of the instability of 
the pion under weak or electromagnetic interactions.
Elaborate extrapolation procedures are used for the extraction of the
scattering lengths from high-energy data.
\par

\subsection{Adler-Weisberger relation} \lb{ss85}

The starting point is the same as for the calculation of the
scattering lengths [Sect. \rf{ss84}]. One chooses for 
$|\ \alpha>$ and $|\ \beta>$ in the WTI nucleon states and for $O$ 
and $j_{5\mu}^A$ axial-vector currents with the pion quantum numbers
\cite{Weisberger:1965hp,Adler:1965ka}. The corresponding two momenta 
$q$ and $k$ are taken to zero. In this limit, there is in addition 
a nucleon pole that contributes, yielding as a residue the 
axial-vector form factor of the nucleon at zero momentum transfer. 
At the end of the operations, one obtains the isospin antisymmetric 
part of the pion-nucleon scattering amplitude at an unphysical point. 
The latter is reexpressed by means of a dispersion relation in terms 
of an integral over physical pion-nucleon cross sections. The final 
formula is
\be \lb{e60}
g_A^2=1-\frac{2F_{\pi}^2}{\pi}\int_{\nu_0}^{\infty}\ \frac{d\nu}{\nu}\
\Big[\sigma^{\pi^-p}(\nu)-\sigma^{\pi^+p}(\nu)\Big],\ \ \ \ \ \ 
\nu=q.p\ .
\ee
The right-hand side yields for $g_A$ the value 1.24, to be
compared with its experimental value 1.27.
\par

\subsection{Gell-Mann--Oakes--Renner formulas} \lb{ss86}

Choose in the WTI [Eq. (\rf{e54})] $|\ \alpha>=|\ \beta>=|0>$ 
and for $O$ the divergence of an axial-vector current 
\cite{GellMann:1968rz}. One has 
\be \lb{61}
\partial^{\nu}j_{5\nu}^B=i\overline\psi_a
\{\mathcal{M},T^B\}_{\ b}^a\gamma_5\psi^b.
\ee
The operators $v^B=-i\overline\psi_a(T^B)_{\ b}^a 
\gamma_5\psi^b$ define the \textit{pseudoscalar densities}.
They transform, in the chiral limit, under the action of the
axial charges as
\be \lb{e62}
[Q_5^A,v^B]=id_{ABC}^{}u^C+i\frac{1}{3}\delta_{AB}u^0,
\ee
where the $u$s are the \textit{scalar densities}
\be \lb{e63}
u^C=\overline\psi_a(T^C)_{\ b}^a\psi^b,\ \ \ \ \
C=1,\ldots,8,\ \ \ \ \ u^0=\overline\psi_a \psi^a, 
\ee
and the coefficients $d$ are fully symmetric in their
indices; they result from the anticommutators of the matrices 
$T$: $\{T^A,T^B\}=d_{ABC}T^C+\frac{1}{3}\delta_{AB}{\mathbf{1}}$. 
\par
The WTI takes the form
\bea \lb{e64}
& &-iq^{\mu}\int dx e^{iq.x} <0|T\ j_{5\mu}^A(x)\ 
\partial^{\nu}j_{5\nu}^B\ |0>\ =\ 
\int dx e^{iq.x} <0|T\ \partial^{\mu}j_{5\mu}^A(x)\ 
\partial^{\nu}j_{5\nu}^B\ |0>\nonumber \\ 
& &\ \ \ \ \ \ \ \ \ \ +\ \int dx e^{iq.x} \delta(x^0) 
<0|\ [j_{50}^A(x),\partial^{\nu}j_{5\nu}^B]\ |0>.
\eea
\par
Intermediate states are only pseudoscalar mesons. In the limit
$q=0$ there are no poles to contribute in the left-hand side and
the latter vanishes, yielding 
\be \lb{e65}
\delta_{AB}\Big\{\ M_{P^A}^2F_{P^A}^2+\sum_{n=1}^{\infty}
M_{P^{nA}}^2F_{P^{nA}}^2\ \Big\}\ =\ -\mathrm{tr}
\{T^A,\{\mathcal{M},T^B\}\}\ <0|\frac{1}{3}u^0|0>.
\ee
From Eqs. (\rf{e43}) and (\rf{e44}) we deduce that 
$F_{P^{An}}^2\ =\ O(\mathcal{M}^2)$, which is much smaller
than $O(\mathcal{M})$ terms. Keeping only $O(\mathcal{M})$ 
quantities, one obtains: 
\be \lb{e66}
\delta_{AB}M_{P^A}^2F_{P0}^2=-\frac{}{}\mathrm{tr}
\{T^A,\{\mathcal{M},T^B\}\}<0|\frac{1}{3}u^0|0>.
\ee
\par
The matrix $\mathcal{M}$ is decomposed in the following way along
the matrices $T$:
$\mathcal{M}=(m_u-m_d)T^3-\frac{1}{\sqrt{3}}(2m_s-m_u-m_d)T^8
+\frac{1}{3}(m_u+m_d+m_s)\mathbf{1}$; the relation 
tr$(T^AT^B)=\frac{1}{2}\delta_{AB}$ fixes the normalization of the 
matrices $T$. Defining   
\be \lb{e67}
B=-\frac{1}{3F_{P0}^2}<0|u^0|0>,\ \ \ \ \ \ \ 
\hat m=\frac{1}{2}(m_u+m_d),
\ee
and neglecting electromagnetism and $\pi^0-\eta-\eta'$ mixings,
Eq. (\rf{e66}) decomposes into the following equations:
\be \lb{e68}
M_{\pi^{+}}^2=M_{\pi^{0}}^2=2\hat mB,\ \ 
M_{K^{+}}^2=(m_s+m_u)B,\ \ M_{K^{0}}^2=(m_s+m_d)B,\ \
M_{\eta}^2=\frac{2}{3}(2m_s+\hat m)B.
\ee
One also verifies here the $SU(3)_V$ Gell-Mann--Okubo formula
(in the isospin limit):
\be \lb{e69}
4M_{K}^2-3M_{\eta}^2-M_{\pi}^2=0.
\ee
\par
The masses $M_P^{}$ and decay couplings $F_P^{}$ are physical 
quantities; threfore, they do not depend on renormalization mass 
scales. However, the quark masses and the scalar densities are 
renormalized under interaction and depend on the renormalization 
mass scale, although their product does not. One must specify, 
when providing values for the quark masses, at which scale 
they have been evaluated. (Usually, they are chosen at a mass scale 
$\mu=2$ GeV.) Also the ratios of quark masses are renormalization 
group invariant.
\par
Numerically, one finds from the above formulas
\be \lb{e70}
\frac{m_s}{\hat m}=26.0,\ \ \ \ \frac{m_u}{m_d}=0.65,\ \ \ \ \ 
\frac{m_s}{m_d}=21.5. 
\ee
One deduces that $m_d>m_u$ and $m_s\gg \hat m$ (cf. also Eq. 
(\rf{e18})). More precise results are obtained by incorporating 
electromagnetic effects \cite{Gasser:1982ap,Leutwyler:1996qg}.
\par
We consider now the vacuum expectation value of the scalar density
$u^0$: $<0|u^0|0>$. With respect to flavor $SU(3)$,
$u^0$ is a singlet operator. However, with respect to
the chiral group $SU(3)_L^{}\otimes SU(3)_R^{}$, it has a more
complicated structure. To display it, we introduce left-handed and 
right-handed quark and antiquark fields:
\be \lb{e71}
\psi_L^a=\frac{1}{2}(1-\gamma_5)\psi^a,\ \ 
\psi_R^a=\frac{1}{2}(1+\gamma_5)\psi^a,\ \  
\overline \psi_{La}=\frac{1}{2}\overline \psi_a(1+\gamma_5),\ \ 
\overline \psi_{Ra}=\frac{1}{2}\overline \psi_a(1-\gamma_5).
\ee
In terms of them, $u^0$ is expressed as
\be \lb{e72}
u^0=\overline \psi_a\psi^a=\overline \psi_{Ra}\psi_L^a
+\overline \psi_{La}\psi_R^a.
\ee
The left-handed and right-handed fields belong to representations of 
different groups, $SU(3)_L^{}$ and $SU(3)_R^{}$, respectively. 
One finds that $u^0$ belongs to the 
$(\overline 3_L,3_R)+(3_L,\overline 3_R)$ representation 
of the chiral group. This is not the singlet representation.
If the vacuum were invariant under chiral transformations, then
$<0|u^0|0>$ would be zero, according to the Wigner-Eckart 
theorem. Its nonvanishing is a sign that the vacuum is not invariant
under chiral transformations and therefore chiral symmetry is 
spontaneously broken. In this case, $<0|u^0|0>$ represents an
\textit{order parameter} of spontaneous chiral symmetry breaking.
One has here an analogy with \textit{ferromagnetism}, where
the magnetization of the atoms has a nonzero value, resulting
from the alignment of their spins along a particular direction,
thus breaking the symmetry of the ground state, which corresponds
to the disordered situation, where the spins are aligned randomly.   
\par

\section{Chiral perturbation theory} \lb{s9}

After the successes of the predictions of low energy theorems, obtained
in the chiral limit or in leading order of explicit chiral symmetry
breaking, one naturally is interested by the calculation of corrective
terms to the leading-order quantities.
Essentially, there are two types of correction that arise:
1) quark mass term contributions; 2) many-Goldstone boson state 
contributions. The latter do not contribute at leading order because 
of damping factors coming from phase space coefficients. At nonleading 
orders, they are no longer negligible. They produce \textit{unitarity 
cuts} with corresponding logarithmic functions of the momenta and the 
masses. Early calculations have been done by Li and Pagels 
\cite{Li:1971vr,Pagels:1974se}.
\par
Calculation of the nonleading chiral symmetry corrections through a 
systematic perturbative approach  has been proposed by 
Glashow and Weinberg \cite{Glashow:1967rx} and by Dashen 
\cite{Dashen:1969eg}.
Initial calculations have been done with the use of the Ward-Takahashi
identities \cite{Dashen:1969bh}. But at higher-orders, this method 
becomes rapidly very complicated.
\par
In this context, Weinberg has made several observations 
\cite{Weinberg:1968de,Weinberg:1969hw}.
1) In spite of the fact that we are in the domain of strong 
interactions, the couplings of the Goldstone bosons with other 
particles and with themselves have turned out to be relatively 
\textit{weak} in the final results of low-energy theorems. 
2) These couplings are reminiscent of \textit{derivative coupling 
types}.
3) All results of low-energy theorems could have been obtained (often
more easily) by using phenomenological chiral invariant Lagrangians 
involving only the hadrons entering into the concerned processes; it 
would be sufficient to do the calculations at the tree level.  
Therefore, nonleading contributions to the low-energy theorems could
be evaluated by calculating loop diagrams with the same Lagrangians.
\par
How to construct chiral invariant hadronic Lagrangians?
For this, one should know the chiral transformation properties of 
the hadronic fields and in particular of the Goldstone boson fields.
\par
The problem of the realization of chiral symmetry through hadronic 
Lagrangians has been systematically studied first by Weinberg
\cite{Weinberg:1968de,Weinberg:1969hw}.
Considering the case of $SU(2)_L^{}\otimes SU(2)_R^{}$ symmetry and 
denoting $\pi^A$ ($A=1,2,3$) the pion fields, which belong to the 
triplet representation of the flavor group (the isospin), one has
\be \lb{e73}
[Q^A,\pi^B]=i\epsilon_{ABC}^{}\pi^C,
\ee
while the transformation rule under the axial charges is left in
a general form:
\be \lb{e74}
[Q_5^A,\pi^B]=-i\Big(\delta_{AB}^{}f(\mbox{\boldmath$\pi$}^2)
+\pi^A\pi^Bg(\mbox{\boldmath$\pi$}^2)\Big).
\ee
($\epsilon_{ABC}^{}=f_{ABC}^{}$ for $A,B,C=1,2,3$.)
Conditions are obtained on the functions $f$ and $g$ from the
action of the chiral algebra [Eqs. (\rf{e25})] on the above
equations and the use of the Jacobi identities. Equations are
obtained for $f$ and $g$ that can be solved. The solution is not 
unique, due to the fact that any pseudoscalar field, having the 
same quantum numbers as the pion, can be considered as 
representing the pion field. These various definitions are related 
to each other by field transformations, which should not have any 
incidence on physical quantities. The general result is that the 
functions $f$ and/or $g$ are nonlinear functions of the pion fields. 
The chiral transformation rules of the pion fields are therefore 
\textit{nonlinear}.
\par
Similarly, one obtains the chiral transformation properties of 
massive hadronic states. The latter transform \textit{linearly}, 
but the transformation coefficients are \textit{nonlinear} functions 
of the pion fields.
\par
To construct chiral invariant Lagrangians, one needs the introduction 
of \textit{chiral covariant derivatives}, in which the role of 
connections is played by pion field dependent terms proportional to 
the derivatives of the pion fields.
One thus finds deductively the property that the couplings of the 
Goldstone boson fields are of the derivative type. 
\par  
Another approach to the construction of chiral invariant Lagrangians
has been developed by Callan, Coleman, Wess and Zumino
\cite{Coleman:1969sm,Callan:1969sn}, which also is 
more easily applicable to the $SU(3)_L^{}\otimes SU(3)_R^{}$ symmetry
case. Instead of considering the detailed form of the transformation 
of the Goldstone fields with respect to the action of the generators
of the group, one here directly considers the transformation rules
at the group element level. For this, one defines the composite
field
\be \lb{e75}
U(x)=e^{{\displaystyle 2i\phi^A(x)T^A/F_{P0}}},
\ee
where $\phi^A$ represent the Goldstone boson fields ($A=1,\ldots,8$),
$T^A$ the matrix representatives of the flavor group generators in the 
`fundamental representation and $F_{P0}$ the Goldstone boson decay 
constants in the chiral limit.
This object is recognized as an element of the 
$SU(3)_L^{}\otimes SU(3)_R^{}$ group, obtained from the application of 
the group operators on the identity element with parameters identified,
up to multiplicative factors, with the Goldstone boson fields. The 
ordinary elements of the group are also of similar type, with parameters 
being now c-numbers. One has the transformation law
\be \lb{e76}
U(x)\ \longrightarrow\ \Omega_R^{}(x)\ U(x)\ \Omega_L^+(x),
\ee
where $\Omega_R$ and $\Omega_L$ are elements of the groups 
$SU(3)_R^{}$ and $SU(3)_L^{}$, respectively. 
(A detailed derivation of Eq. (\rf{e76}) can be found in Refs.
\cite{Manohar:1996cq,Kubis:2007iy}.) They are considered
as $x$-dependent to allow for the introduction of covariant derivatives.
The latter are defined from Eq. (\rf{e76}) with external currents or
sources playing the role of connections. The chiral invariant 
Lagrangian is constructed with the aid of the covariant derivatives,
also including couplings to external scalar and pseudoscalar densities.
This method  of approach leads naturally, by means of the external
sources, to the definitions of Green's functions of currents and 
densities. It has been adopted by Gasser and Leutwyler in their 
formulation of the chiral effective field theory of Goldstone bosons 
in the framework of QCD \cite{Gasser:1983yg,Gasser:1984gg}.
\par  
Explicit chiral symmetry breaking is further introduced through mass 
terms of Goldstone bosons \cite{Weinberg:1968de,Weinberg:1969hw} or 
through external sources representing quark masses
\cite{Gasser:1983yg,Gasser:1984gg}. 
\par
Next, one calculates loop corrections. Since Goldstone boson couplings 
are of the derivative type, each loop introduces, through its vertices,
additional powers of the Goldstone boson momenta. At low energies, the
latter are small, of the same order of magnitude as the Goldstone
boson masses. Increasing the number of loops increases in turn the 
powers of the momenta and hence decreases the magnitude of the terms 
in comparison with the leading ones \cite{Weinberg:1978kz}.
\par
Therefore, one is led to a systematic power counting rule. The
Goldstone boson masses squared $M_{P}^2$ (or equivalently the quark
masses $\mathcal{M}$) and momenta squared, $q^2,\ k^2$, etc., are 
considered as perturbation theory parameters. Their higher powers will 
correspond to higher-order terms. Thus, the first-order corrections 
will involve at most one-loop diagrams. Chiral perturbation theory 
involves therefore expansions in quark masses and Goldstone boson 
momenta, rather than in coupling constants.
\par     
A crucial point that remains to be dealt with is {renormalization}.
In order for the phenomenological hadronic Lagrangian to describe
correctly the underlying theory (QCD), it is necessary to consider 
\textit{the most general} chiral invariant Lagrangian together with
all possible mass terms . Therefore, the latter will involve an 
infinite series of terms with an increasing number of derivatives and 
powers of the masses. It will involve, from the start 
an infinite number of unknown constants. However, the Lagrangian will be 
ordered according to the number of derivatives or powers of masses it 
involves. 
For instance, for the purely Goldstone boson Lagrangian one has the
expansion \cite{Gasser:1983yg,Gasser:1984gg}
\be \lb{77}
\mathcal{L}=\mathcal{L}_2^{}+\mathcal{L}_4^{}+\cdots 
+\mathcal{L}_{2n}^{}+\cdots \ ,
\ee
where the generic subscript $2n$ designates the total number of the
derivatives and of the power of the Goldstone boson masses contained 
in the related term. At each finite order of the expansion, there 
corresponds a finite number of independent terms. The 
leading-order contributions come from $\mathcal{L}_2^{}$, those of 
the next-to-leading order from $\mathcal{L}_2^{}+\mathcal{L}_4^{}$, etc.
\par
Calculation of loops introduces divergences. These are then absorbed
by the renormalization of the coupling constants of the higher-order 
parts of the Lagrangian. These coupling constants are called
\textit{low energy constants (LEC)}. They also contain information about 
the contributions of the high-mass particles which do not appear explicitly
in the Lagrangian.
Thus, in the Lagrangian describing $\pi-\pi$ interaction,
the $\rho$-meson is absent, but since the Lagrangian 
is describing only low-energy regions, the $\rho$-meson
field is in some sense integrated out from the total Lagrangian of
the theory and its effects are enclosed in the resulting LECs and
the accompanying expressions.
\par
At each finite order of the perturbation theory, there is only a
finite number of LECs, which should be determined from
experiment. Then the theory becomes predictive at that order.
\par       
The phenomenological chiral hadronic Lagrangians that are constructed
from the previous principles define \textit{the chiral effective field
theories} designed to describe definite sectors of hadrons.
Applications have concerned processes involving $\pi,\ K\ ,\eta$  
mesons \cite{Gasser:1984ux,Gasser:1984pr,Bernard:1990kw,Colangelo:2001df,
Bijnens:2006zp}, leading to improved precision with respect to the 
low-energy theorems. The method has been generalized to include 
baryon-Goldstone boson interactions \cite{Gasser:1987rb,Jenkins:1990jv,
Burdman:1992gh,Wise:1992hn,Yan:1992gz,Bernard:1992qa,Cho:1992cf,
Meissner:1997ws,Becher:1999he}.
\par
In conclusion, the chiral effective field theory approach enables one 
to convert, at low energies and in the hadronic world, the highly 
complicated nonperturbative structure of QCD into a more transparent
and familiar framework, with the sole aid of the symmetry properties
of the theory.
\par  

\section{Anomalies} \lb{s10}

Invariance properties of Lagrangians established at the classical 
level may sometimes not survive quantization. In QFT, the 
renormalization process deals with divergent quantities, and some
conventional operations, like interchange of limits or translation of
integration variables, may not be legitimate.
\par
That question has been acutely raised with the study of the decay process
of the neutral pion into two photons ($\pi^0\rightarrow\gamma\gamma$).
Using the usual Ward-Takahashi identities (WTI) of chiral symmetry, one 
derives a low-energy theorem stating that the corresponding decay width 
should vanish in the chiral limit, while, on experimental grounds, it 
is far from vanishing or being small. This paradox has been 
analyzed by Adler and Bell and Jackiw 
\cite{Adler:1969gk,Bell:1969ts}, who have shown that in QED with 
massless fermions, coupled to external axial-vector currents and 
pseudoscalar densities, triangle diagrams, corresponding to fermion 
loops with two vector vertices and one axial-vector vertex, violate, 
through the WTI, the conservation of the axial-vector current. 
This result is unchanged, and more precisely unrenormalized, by 
higher-order radiative corrections \cite{Adler:1969er}. 
Adler has proposed to include the above anomalous contribution in the 
divergence of the axial-vector current, which now reads, for massive
fermions, \cite{Adler:1969gk} 
\be \lb{78}
\partial^{\mu}j_{5\mu}=2im\overline\psi\gamma_5\psi
+\frac{e^2}{16\pi^2}\epsilon^{\mu\nu\lambda\sigma}F_{\mu\nu}^{}
F_{\lambda\sigma}^{},
\ee
where $F$ is the electromagnetic field strength and $\epsilon$ the
totally antisymmetric tensor ($\epsilon^{0123}=-\epsilon_{0123}^{}=-1$). 
Thus, in the presence of electromagnetism, the axial symmetry is 
explicitly broken, irrespective of the fermion masses, a result not 
evident at the classical level.
\par
From a more general standpoint, the appearance of anomalies can
be understood as coming from the necessity of renormalizing the 
fermionic determinant in the path integral formalism. There are
no regularization schemes that preserve chiral invariance
\cite{Fujikawa:1979ay}.
\par  
The general structure of anomalies in the case of non-Abelian currents
has been studied by Bardeen and Wess and Zumino
\cite{Bardeen:1969md,Wess:1971yu}. The anomalous 
contributions appear through the presence of an odd number of 
axial-vector currents at the vertices of fermion one-loop diagrams, 
the number of vertices varying from three to five. Considering the 
case of one axial-vector current in the presence of other vector 
currents, the expression of its divergence becomes
\bea \lb{79}
\partial^{\mu}j_{5\mu}^A&=&i\overline\psi\{\mathcal{M},T^A\} 
\gamma_5\psi
+\frac{N_c}{16\pi^2}\epsilon^{\mu\nu\lambda\sigma}\
\mathrm{tr_f}(T^AT^BT^C)\ F_{\mu\nu}^{B}F_{\lambda\sigma}^{C}
\nonumber\\
& &\ \ \ \ \ +\frac{g^2N_f}{32\pi^2}\epsilon^{\mu\nu\lambda\sigma}
\delta_{A0}\sum_{a=1}^{N_c^2-1}G_{\mu\nu}^{a}G_{\lambda\sigma}^{a},
\ \ \ \ \ A=0,1,\ldots,N_f^2-1,
\eea
where the $F$s are non-Abelian color singlet field strengths 
of external vector currents, playing the role of sources, the $G$s 
the gluon field strengths, $N_f$ and $N_c$ the numbers of quark 
flavors and colors, respectively, $\mathrm{tr_f}$ the trace in flavor 
space, and we have also included in the formula the contribtion of 
flavor singlet currents, with the convention $T^0=\mathbf{1}$.
\par
We note that in the flavor singlet case, the anomaly receives also a 
contribution from the gluon fields. The latter is a dimension-four
operator and, contrary to the usual mass terms, its effect is hard 
and could not be treated in general as a perturbation (cf. the comments 
after Eq. (\rf{e14})). The presence of the anomaly in the divergence 
of the singlet axial-vector current explains the large mass value of 
the $\eta'$ meson, which cannot be considered, in the chiral limit,
as a Goldstone boson \cite{'tHooft:1986nc}.  
\par 
The presence of anomalies destroys the renormalizability and unitarity
properties of field theories. This is not the case of QCD itself, since 
it does not contain axial-vector couplings; the anomalies may appear here 
only when external currents are considered. However, the situation 
is different with the Standard Model, which contains axial-vector 
couplings through its weak interaction part. A general theorem states 
that the anomalies will intrinsically disappear if the sum of the 
charges of all fermions of the theory is equal to zero. This is
indeed the case of the Standard Model, where the sum of the charges 
of the quarks and of the leptons cancel to zero, thus confirming the 
consistency of the theory 
\cite{Itzykson:1980rh,Kaku:1993ym,Bouchiat:1972iq}.
\par
Apart from their implication in filling a hole in the chiral
Ward-Takahashi identities, the anomalies play also an important
role in the analysis of the phases of a theory. Such an attempt has
been undertaken by 't Hooft with the concept of 'naturalness'
\cite{'tHooft:1979bh}.
If physics is described by different effective theories at different
scales, the anomaly structure should be preserved from one scale to 
the other when the transition concerns only one part of the particles.
(This is a consequence of the zero-anomaly condition for the global
theory.) This requirement can be applied in particular to QCD. The
latter is described at high energies by elementary quark and gluon
fields, while at low energies it is described by hadrons, leptons 
remaining the same. As a consequence of the global zero-anomaly 
condition, anomalies resulting from both descriptions should match 
each other. This is not a trivial requirement, since quarks belong 
to the triplet representation of the flavor group (in the case of 
$SU(3)_V^{}$), while hadrons belong to higher representations, 
like octets, decuplets, etc. The net result of the investigation shows 
that when the number of quark flavors is greater than two, only the phase 
of spontaneous chiral symmetry breakdown can fulfill the anomaly 
matching condition \cite{'tHooft:1979bh,Frishman:1980dq,Coleman:1982yg,
Coleman:1980mx}. The above analysis brings therefore an 
indirect theoretical proof of spontaneous chiral symmetry breaking in 
QCD with more than two quark flavors. This result has been completed
with the proof of Vafa and Witten that in vector-like gauge theories, 
such as QCD, the vector flavor group and parity cannot be spontaneously 
broken \cite{Vafa:1983tf,Vafa:1984xg}.
\par    

\vspace{0.5 cm}
\noindent
\textbf{Acknowledgments}. I thank the Organizing Committee of the
Conference Quark Confinement and the Hadron Spectrum XII for 
inviting me to give this lecture. I also thank V\'eronique Bernard 
and Bachir Moussallam for discussions and useful information.
\par


\begin{thebibliography}{75}

\bibitem{Coleman:1985}S. Coleman, \textit{Aspects of symmetry}
(Cambridge University Press, Cambridge, UK, 1985).
\bibitem{Itzykson:1980rh}C. Itzykson and J.-B. Zuber,
\textit{Quantum field theory} (McGraw-Hill, New York, 1980).  
\bibitem{Kaku:1993ym}M. Kaku, \textit{Quantum field theory: 
A modern introduction} (Oxford University Press, New York, USA, 
1993).
\bibitem{Peskin:1982mu}M. E. Peskin, Les Houches Lectures, 1982,
SLAC-PUB-3021.
\bibitem{Gasser:1982ap}J. Gasser and H. Leutwyler, \prp \textbf{87},
77 (1982).
\bibitem{Manohar:1996cq}A. V. Manohar, Lect. Notes Phys. \textbf{479},
311 (1997).
\bibitem{Scherer:2002tk}S. Scherer, Adv. Nucl. Phys. \textbf{27}, 277
(2003).
\bibitem{Gasser:2003cg}J. Gasser, Lect. Notes Phys. \textbf{629}, 1
(2004).
\bibitem{Ecker:2006xd}G. Ecker, arXiv: hep-ph/0604165. 
\bibitem{Bernard:2006gx}V. Bernard and U.-G. Meissner, Ann. Rev.
Nucl. Part. Phys. \textbf{57}, 33 (2007).
\bibitem{Kubis:2007iy}B. Kubis, arXiv: hep-ph/0703274.
\bibitem{Brambilla:2014jmp}N. Brambilla \textit{et al.}, \epj C
\textbf{74:2981}, 1 (2014).
\bibitem{Preparata:1969hg}G. Preparata and W. I. Weisberger, \pr
\textbf{175}, 1965 (1968).
\bibitem{Goldstone:1961eq}J. Goldstone, \nc \textbf{19}, 154 (1961).
\bibitem{Goldstone:1962es}J. Goldstone, A. Salam and S. Weinberg,
\pr \textbf{127}, 965 (1962).
\bibitem{Nambu:1960xd}Y. Nambu, \prl \textbf{4}, 380 (1960).
\bibitem{Nambu:1961tp}Y. Nambu and G. Jona-Lasinio, \pr \textbf{122},
345 (1961).
\bibitem{GellMann:1960np}M. Gell-Mann and M. L\'evy, \nc \textbf{16},
705 (1960).
\bibitem{Agashe:2014kda}K. A. Olive \textit{et al.}, Particle Data
Group, \cpc \textbf{38}, 090001 (2014).
\bibitem{Goldberger:1958vp}M. L. Goldberger and S. B. Treiman,
\pr \textbf{111}, 354 (1958).
\bibitem{Baru:2010xn}V. Baru \textit{et al.}, \plb \textbf{694},
473 (2011).
\bibitem{Ward:1950xp}J. C. Ward, \pr \textbf{78}, 182 (1950).
\bibitem{Takahashi:1957xn}Y. Takahashi, \nc \textbf{6}, 371 (1957).
\bibitem{Callan:1966hu}C. G. Callan and S. B. Treiman, \prl 
\textbf{16}, 153 (1966).
\bibitem{Weinberg:1966kf}S. Weinberg, \prl \textbf{17}, 616 (1966).
\bibitem{Tomozawa:1966jm}Y. Tomozawa, \nc A \textbf{46}, 707 (1966).
\bibitem{Weisberger:1965hp}W. I. Weisberger, \prl \textbf{14}, 1047
(1965).
\bibitem{Adler:1965ka}S. L. Adler, \prl \textbf{14}, 1051 (1965).
\bibitem{GellMann:1968rz}M. Gell-Mann, R. J. Oakes and B. Renner,
\pr \textbf{175}, 2195 (1968).
\bibitem{Leutwyler:1996qg}H. Leutwyler, \plb \textbf{378}, 313 (1996).
\bibitem{Li:1971vr}L.-F. Li and H. Pagels, \prl \textbf{26}, 1204
(1971).
\bibitem{Pagels:1974se}H. Pagels, \prp \textbf{16}, 219 (1975).
\bibitem{Glashow:1967rx}S. L. Glashow and S. Weinberg, \prl
\textbf{20}, 224 (1968).
\bibitem{Dashen:1969eg}R. F. Dashen, \pr \textbf{183}, 1245 (1969).
\bibitem{Dashen:1969bh}R. F. Dashen and M. Weinstein, \prl \textbf{22},
1337 (1969).
\bibitem{Weinberg:1968de}S. Weinberg, \pr \textbf{166}, 1568 (1968).  
\bibitem{Weinberg:1969hw}S. Weinberg, \pr \textbf{177},2604 (1969).
\bibitem{Coleman:1969sm}S. R. Coleman, J. Wess and B. Zumino, \pr
\textbf{177}, 2239 (1969).
\bibitem{Callan:1969sn}C. G. Callan, Jr., S. R. Coleman, J. Wess
and B. Zumino, \pr \textbf{177}, 2247 (1969).
\bibitem{Gasser:1983yg}J. Gasser and H. Leutwyler, \ap \textbf{158},
142 (1984).
\bibitem{Gasser:1984gg}J. Gasser and H. Leutwyler, \npb \textbf{250},
465 (1985).
\bibitem{Weinberg:1978kz}S. Weinberg, Physica \textbf{A96}. 327 (1979).
\bibitem{Gasser:1984ux}J. Gasser and H. Leutwyler, \npb \textbf{250},
517 (1985).
\bibitem{Gasser:1984pr}J. Gasser and H. Leutwyler, \npb \textbf{250},
539 (1985).
\bibitem{Bernard:1990kw}V. Bernard, N. Kaiser and U.-G. Meissner,
\npb \textbf{357}, 129 (1991).
\bibitem{Colangelo:2001df}G. Colangelo, J. Gasser and H. Leutwyler,
\npb \textbf{603}, 125 (2001).
\bibitem{Bijnens:2006zp}J. Bijnens, Prog. Part. Nucl. Phys. \textbf{58},
521 (2007).
\bibitem{Gasser:1987rb}J. Gasser, M. E. Sainio and A. \v{S}varc, \npb
\textbf{307}, 779 (1988).
\bibitem{Jenkins:1990jv}E. E. Jenkins and A. V. Manohar, \plb 
\textbf{255}, 558 (1991).
\bibitem{Burdman:1992gh}G. Burdman and J. F. Donoghue, \plb \textbf{280},
287 (1992).
\bibitem{Wise:1992hn}M. B. Wise, \prd \textbf{45}, R2188 (1992).
\bibitem{Yan:1992gz}T.-M. Yan \textit{et al.}, \prd \textbf{46}, 1148
(1992).
\bibitem{Bernard:1992qa}V. Bernard, N. Kaiser, J. Kambor and 
U.-G. Meissner, \npb \textbf{388}, 315 (1992).
\bibitem{Cho:1992cf}P. L. Cho, \npb \textbf{396}, 183 (1993).
\bibitem{Meissner:1997ws}U.-G. Meissner, arXiv: hep-ph/9711365.
\bibitem{Becher:1999he}T. Becher and H. Leutwyler, \epj C \textbf{9},
643 (1999).
\bibitem{Adler:1969gk}S. L. Adler, \pr \textbf{177}, 2426 (1969).
\bibitem{Bell:1969ts}J. S. Bell and R. Jackiw, \nc \textbf{A60}, 47
(1969).
\bibitem{Adler:1969er}S. L. Adler and W. A. Bardeen, \pr \textbf{182},
1517 (1969).
\bibitem{Fujikawa:1979ay}K. Fujikawa, \prl \textbf{42}, 1195 (1979).
\bibitem{Bardeen:1969md}W. A. Bardeen, \pr \textbf{184}, 1848 (1969).
\bibitem{Wess:1971yu}J. Wess and B. Zumino, \plb \textbf{37}, 95 
(1971). 
\bibitem{'tHooft:1986nc}G. 't Hooft, \prp \textbf{146}, 357 (1986).
\bibitem{Bouchiat:1972iq}C. Bouchiat, J. Iliopoulos and P. Meyer,
\plb \textbf{38}, 519 (1972).
\bibitem{'tHooft:1979bh}G. 't Hooft, NATO Sci. Ser. B \textbf{59}, 
135 (1980).
\bibitem{Frishman:1980dq}Y. Frishman, A. Schwimmer, T. Banks and
S. Yankielowicz, \npb \textbf{177}, 157 (1981).
\bibitem{Coleman:1982yg}S. R. Coleman and B. Grossman, \npb 
\textbf{203}, 205 (1982).
\bibitem{Coleman:1980mx}S. R. Coleman and E. Witten, \prl \textbf{45},
100 (1980).
\bibitem{Vafa:1983tf}C. Vafa and E. Witten, \npb \textbf{234}, 173
(1984).
\bibitem{Vafa:1984xg}C. Vafa and E. Witten, \prl \textbf{53}, 535
(1984).

\end{thebibliography}
\end{document}